\documentclass[%
 reprint,
 amsmath,amssymb,
 aps,
]{revtex4-2}

\usepackage{graphicx}
\usepackage{dcolumn}
\usepackage{bm}
\usepackage{mathrsfs}
\usepackage{float}      
\usepackage[table]{xcolor}      
\usepackage{subcaption} 
  
\usepackage{multirow}
\usepackage{makecell}
 \usepackage{amsmath}
\usepackage{graphicx}
\usepackage{url}
\usepackage{fancyhdr}
\usepackage{footmisc}
\usepackage{algorithmic}
\usepackage{listings}
\usepackage{fancyvrb}
\usepackage{graphicx}
\usepackage{hyperref}
 
\begin{document}

\graphicspath{ {./} }
\preprint{APS/123-QED}

\title{Concurrence and entanglement on a 16-site spin-1/2 pyrochlore cluster
}

\author{C.~Wei}
\email{cw7734@mun.ca}
\author{S.~H.~Curnoe}%
 \email{curnoe@mun.ca}
\affiliation{%
 Department of Physics and Physical Oceanography, Memorial University of Newfoundland,
St. John's, Newfoundland \& Labrador, Canada A1B 3X7
}%

\date{\today}

\begin{abstract}
We examine the entanglement of the ground state of a 16-site spin-1/2 pyrochlore cluster in the quantum spin ice regime via various calculations of 
${\cal I}$-concurrence. Exact ground state solutions to a quantum spin Hamiltonian with four nearest-neighbour exchange parameters were obtained using exact diagonalization.  We present results for the ground state ${\cal I}$-concurrence in a region within the parameter space of the model where the ground state is a singlet.   We discuss variations in the ${\cal I}$-concurrence in the context of the composition of the ground state and we demonstrate how a lattice distortion disentangles the state.

\end{abstract}

                              
\maketitle

\section{Introduction}\label{sec:Intro}
The concept of quantum entanglement was first introduced in 1935  in a
discussion of the completeness of quantum theory
\cite{Einstein_Podolsky_Rosen_1935}. However, it was not until 1967 that an entangled state of two photons was achieved in experiment \cite{Kocher_Carl_Commins_1967}. 
Qualitatively, entanglement in a pure quantum state means that measurement of one part of the state affects subsequent measurement of the other part. An unentangled state $|\psi\rangle$ is {\em separable}, that is, it can be expressed as a single product over its parts, $|\Psi\rangle = |\psi\rangle \otimes|\phi\rangle \otimes |\xi\rangle \ldots$.
A bipartitite two-dimensional system (for example, two photons or two spins) in the state
$\alpha|++\rangle + \beta|+-\rangle + \gamma|-+\rangle + \delta|--\rangle$
is separable if the quantity $e = |\alpha\delta - \beta\gamma| = 0$.
Thus the simplest way to quantify the entanglement in this kind of system is to evaluate
$e$:  $e$ equals 0 for an 
unentangled state and has a maximum value of $1/2$, corresponding to a maximally entangled state;
other well-known entanglement measures can be expressed in terms of $e$: the {\em entanglement entropy}, is a monotonically increasing function of $e$, while the {\em concurrence} is simply $2e$.  As we shall discuss, the concurrence is a particularly useful construction as it naturally lends itself to generalizations for multi-partite or higher dimensional systems.



In this paper, we examine entanglement in the ground of a frustrated quantum spin system, specifically, a pyrochlore ferromagnet with anisotropic nearest neighbour exchange interactions, whose ground state is predominately a superposition of classical spin ice states.  In pyrochlore magnets the spins are located at the vertices of a network of corner-sharing tetrahedra; a classical spin ice state is one that has exactly two spins pointing into and two spins pointing out of every tetrahedron on the lattice. 

The model that we consider reflects the high symmetry of the lattice: pyrochlore crystals belong to the space group $Fd3\bar{m}$ (No.\ 227 $O_h^7$), which is based on the octahedral point group $O_h$ and face-centred cubic lattice translations.  In some sense, it is high the symmetry of the lattice that gives rise to highly entangled eigenstates.  Every eigenstate belongs to one of the irreducible representations of the symmetry group, in particular it means that singlet states must be superpositions of all kets that are related to each other by the crystal symmtries.  We will also consider the effect of the lowering the lattice symmetry from $Fd3\bar{m}$ to $F\bar{4}3m$. This symmetry breaking is equivalent to introducing a ``breathing mode" on the tetrahedra of the lattice, which alters  interactions between tetrahedra  and tends to distentangle the parts of the state associated with different tetrahedra. 

While entanglement between two-dimensional systems with only two parts is captured by the quantity $e$, different types of concurrence measures are used to quantity different aspects of entanglement in multi-partitite systems, which we describe in Section II.  In Section III, we present the spin Hamiltonian 
for nearest neighbour exchange interactions on the pyrochlore lattice - a model which is expected to host a highly entangled ground state. 
Using the results obtained from exact diagonalization of the spin Hamiltonian, in Section IV we present numerical results for a set of concurrence measures in order to study the ground state entanglement between different parts of our system, from individual spins to the set of four spins on a single tetrahedron. In Section V we discuss variations in the ${\cal I}$ concurrence in the context of the composition of the exact ground state wavefunction and concluding remarks are offered in Section VI.

\section{Concurrence}\label{sec:Model}

There are several different but equivalent procedures to evaluate the concurrence of bipartite, 2D systems \cite{Rungta_Pranaw_2001}\cite{Wootters_William_1998}\cite{Hill_Sam_Wootters_William_1997}, 
however there are two approaches that are  easily generalized to larger systems.  
In the first approach, one begins by writing a state that is two copies of the original bipartite state, 
$|\psi\rangle|\psi\rangle$. Then, the
anti-symmetrization operator $A\otimes A$
is applied to each part, with $A|+-\rangle = |+-\rangle - |-+\rangle$ (note that in these kets both entries are associated with the same part).  In our example state $|\psi\rangle = \alpha|++\rangle+
\beta|+-\rangle + \gamma|-+\rangle + \delta|--\rangle$ this yields
$$A\otimes A|\psi\rangle |\psi\rangle
= (\alpha\delta - \beta\gamma)(|+-\rangle - |-+\rangle)\otimes 
(|+-\rangle - |-+\rangle).
$$
The concurrence is given by 
$$
C(\psi) = 2^{1/2} \sqrt{\langle \psi|\langle \psi|A\otimes A|\psi\rangle|\psi\rangle}
= 2|e|.
$$
For multipartite states with $N$ parts the obvious extension
employs operators $A^{\otimes N}$.  Additionally, operators of the form $A^{\otimes P} \otimes S^{\otimes Q}$ where $P>0$ is even and $P+Q=N$ can provide complementary measures of entanglement, for example, there exists entangled states where the measure based on $A^{\otimes N}$ vanishes but is non-zero for combinations of $A$ and $S$ operators.  The drawback of these measures is that there too many different possibilities to compute and analyze: for $N=16$ we have 36407 concurrence functions, although many of these will be redundant depending on the symmetry of the state.  

The second approach computes the concurrence of the bipartite, two-dimensional state as 
$$
C(\psi) = \sqrt{2(1 - \mbox{Tr} \rho_r^2)}
$$
where $\rho_r$ is the reduced density matrix, obtained by taking the trace over either one of the parts. This quantity equals $2e$, in agreement with the first approach.  For $N$-partite two-dimensional systems, it can be shown that 
\begin{equation}
C(\psi) = 2^{1-N/2}\sqrt{2^N-2
-\sum_i \mbox{Tr} \rho_i^2},
\label{eq:concur2}
\end{equation}
where $\rho_i$ are all possible 
reduced density matrices, is
equivalent to the sum over all 
concurrence measures of the first approach \cite{Li_Fei_Wang_2010}.  The reduced density matrices $\rho_i$ are  obtained by taking the trace of $\rho = |\psi\rangle \langle \psi|$ over one or more of the parts. 
To understand why this
quantity measures entanglement, consider the following: there are
$2^N-2$ terms in the sum over $i$.
If the state is separable,
$|\psi\rangle = |\psi_i\rangle\otimes |\psi_j\rangle$ ({\em i.e.}, the parts $i$ and $j$ are unentangled),
then the trace of $\rho = |\psi\rangle \langle \psi|$ over the $j$th part
yields 
$\rho_i = |\psi_i\rangle\langle \psi_i|$; since this represents a pure state, the trace of $\rho_i^2$ equals one. Otherwise, if the parts $i$ and $j$ entangled, $\rho_i$ will not represent a pure state, so $\mbox{Tr}\rho_i^2 <1$, and there will be a non-zero contribution to $C(\psi)$.



Along the line of the second approach, 
for a bipartite pure state $|\psi\rangle$ in arbitrary dimensions, one can define the ${\mathcal {I}}$-concurrence as
\begin{equation}
{\displaystyle {\mathcal {C}}_{\mathcal {I}}(\psi )={\sqrt {2(1-{\text{Tr}}\rho _{\mathcal {S}}^{2})}}}
\label{eq:Iconcur}
\end{equation}
where $\rho _{\mathcal {S}}$  is the reduced density matrix of the subsystem $\mathcal {S}$
of the pure state $|\psi\rangle$. 
Note that 
$\mbox{Tr} \rho^2_{\cal S} = \mbox{Tr} \rho^2_{\bar{\cal S}}$,  where 
$\bar{\cal S}$ is the complement of ${\cal S}$, so
the result is the same regardless of which part is used as the subsystem. 
${\text{Tr}}\rho _{\mathcal {S}}^{2}$ is minimized when 
$\rho _{\mathcal {S}}$ is a diagonal matrix with elements $1/d$, where $d$ is the dimension of the smaller subsystem; this yields an upper limit on the ${\cal I}$-concurrence of $\displaystyle {\mathcal {C}}_{\mathcal {I}}(\psi )\le \sqrt{2(1-1/d)}$ \cite{florian_andre_carvalho_marek_measure_entagled_states}.
The ${\mathcal {I}}$-concurrence can also be used evaluate the entanglement of $n$ parts of $N$-partitite, 
2D dimensional system with the rest of the parts by considering the system to be bipartite with dimensions 
$2^n$ and $2^{N-n}$.  Note that this is the same as 
considering only one term in the sum over $i$ in Eq.\ 
\ref{eq:concur2}.
In the following, we will apply the ${\mathcal {I}}$-concurrence to evaluate the entanglement between different parts of pure singlet states on  a pyrochlore spin system. 
A singlet ground state occurs for a wide range of the model parameters, and we assume that $T=0$ in order to compute the concurrence.

\section{The Spin Hamiltonian}
\subsection{Nearest-neighbour exchange interaction for the pyrochlore lattice}
In pyrochlore magnets, the spins are located at the vertices of a corner-sharing tetrahedral network. Each spin resides on two tetrahedra and has six nearest neighbours, which are the spins occupying the other vertices of the two tetrahedra.  Therefore each nearest-neighbour interaction corresponds to an edge of a tetrahedron (and vice versa). 
There are four equivalent spin sites (numbered 1 to 4), corresponding to the four vertices of a tetrahedron, with site
symmetry $D_{3d}$, where the three-fold axes points along one of the cube diagonals.  It is convenient to adopt these axes as the spin-quantization axes ({\em i.e.} the $z$-axes). 

We consider a general nearest-neighbor exchange interaction for spin-1/2 spins,
\begin{equation}
    H=\sum_{\langle i,j\rangle}{\cal J}_{i,j}^{\mu \nu}S^\mu_i S^\nu_j,
\end{equation}
where the sum over $\langle i,j\rangle$ runs over pairs of nearest-neighbour spins
and $\vec{S}_i=(S^x_i, S^y_i,S^z_i)$ is the spin operator for the $i$th site. ${\cal J}_{i,j}^{\mu \nu}$ are exchange constants which are constrained by the space group symmetry of the crystal. In pyrochlore magnets
there are only four independent exchange constants allowed by symmetry \cite{Curnoe_physrevb_2008}. 
Therefore it is convenient to express the spin Hamiltonian as
\begin{equation}
H = {\cal J}_{1}X_1+{\cal J}_{2}X_2+{\cal J}_{3}X_3+{\cal J}_{4}X_4,
\label{eq:hamil}
\end{equation}
where ${\cal J}_{a}$ are the exchange constants and
\begin{align*}
&X_1=-\frac{1}{3}\sum_{\langle i,j \rangle}S_{iz}S_{jz}\\
&X_2=-\frac{\sqrt{2}}{3}\sum_{\langle i,j\rangle}[\Lambda_{s_is_j}(S_{iz}S_{j+}+S_{jz}S_{i+})+{\rm h.c.}]\\
&X_3=\frac{1}{3}\sum_{\langle i,j\rangle}[\Lambda_{s_is_j}^*S_{i+}S_{j+}+ {\rm h.c.}]\\
&X_4=-\frac{1}{6}\sum_{\langle i,j \rangle}(S_{i+}S_{j-}+ {\rm h.c.}).
\end{align*}
In these expressions, the spin operators $\vec{S}_i$ are given in terms of a set of local axes, such that the local $z$-axes are the spin quantization axes described above (see Refs.\ \cite{Curnoe2007_PhysRevB_2007, Curnoe_physrevb_2008} for
more details),
 $\Lambda_{ss'}$ are phases which depend on the site numbers: $\Lambda_{12}=\Lambda_{34}=1$ and $\Lambda_{13}=\Lambda_{24}=\Lambda_{14}^{*}=\Lambda_{23}^{*}=\varepsilon\equiv \exp(\frac{2\pi i}{3})$, and $S_{\pm}=S_x\pm i S_y$.
Note that when ${\cal J}_{1}={\cal J}_{2}={\cal J}_{3}={\cal J}_{4} \equiv {\cal J}$ the exchange interaction is simply the isotropic (Heisenberg) exchange, $H_{\rm iso} = {\cal J}\sum\limits_{\langle i,j \rangle}\Vec{J_{i}}\cdot \Vec{J_{j}}$ \cite{Curnoe_physrevb_2008}.

Classical spin ice states are a highly degenerate set of states that obey the ``ice rule:" two spins pointing into and two spins pointing out of every tetrahedron in the lattice.  Such states are eigenstates of the term $X_1$ in the Hamiltonian, and are the ground states of ${\cal J}_1 X_1$ when ${\cal J}_1 < 0$. A quantum spin ice state is predominately a linear superposition of these classical states. 

\subsection{Single tetrahedron
\label{tetra}}
On a single tetrahedron with four spins, there are six classical, degenerate spin ice states, 
$|++--\rangle$, 
$|--++\rangle$, 
$|+--+\rangle$, $|-++-\rangle$,
$|+-+-\rangle$, and $|-+-+\rangle$, where `$+/-$' represents a spin pointing out of/into the tetrahedron. In the quantum case (when terms $X_2$ to $X_4$ are present), symmetry considerations predict that the six-fold degeneracy will be lifted into a singlet, a doublet, and a triplet. The singlet state, denoted $|A_1\rangle$,  is an equally weighted superposition of all six of these states (with no additional phases).  This state is the ground state of the Hamiltonian for a single tetrahedron when ${\cal J}_1 <0$ and ${\cal J}_4 >0$ and
$|{\cal J}_{2,3}| \lessapprox {\cal J}_4$.
Considering one spin as a two-dimensional subsystem, the reduced density of of the state $|A_1\rangle$ is 
 $\rho_{\cal S} = 
\left(\begin{array}{cc}
1/2 & \\
 & 1/2 \end{array} \right)$ and   the ${\cal I}$-concurrence  (\ref{eq:Iconcur})  is
 $C(A_1) = 1$, a maximum.  For two spins comprising a four-dimensional subsystem, we have 
$\rho_{S} = \left(\begin{array}{cccc}
    1/6 \\
     & 1/3 & 1/3   \\
     & 1/3 & 1/3 \\
     & & & 1/6
\end{array}\right)$, yielding an ${\cal I}$-concurrence of 1, somewhat less than the maximimum of $\sqrt{3/2}$, which reflects the degeneracy and constraints imposed by the ice rule: 
 although the state is not separable, some terms can be factored,
 $|A_1\rangle 
 = \frac{1}{\sqrt 6}(|++--\rangle + |--++\rangle + (|+-\rangle + |-+\rangle) \otimes (|+-\rangle + |-+\rangle))$,
 ultimately resulting in a smaller ${\cal I}$-concurrence.

In the 16-spin system (a cube containing four tetrahedra), assuming periodic boundary conditions, there are 90 classical spin ice states. In the quantum system, only 6 are singlets -  the remaining 84 are doubly, triply or six-fold degenerate, and all are superpositions of the classical spin ice states.  In the following, we consider only sets of exchange constants that lead to a singlet ground state, and we assume that $T=0$ so that the system is described by a pure state in order to 
calculate the ${\cal I}$-concurrence~(\ref{eq:Iconcur}).

\subsection{Breathing mode lattice distortion}
We will also consider the effect of breathing-mode lattice distortion.  In the pyrochlore lattice, the corner-sharing tetrahedra alternate between two different orientations (related by 90 degree rotations), which we label $A$ and $B$.  Since each nearest neighbour interaction is associated is an edge of a tetrahedron, we can express the Hamiltonian as $H = H_A + H_B$, where $H_A$ and $H_B$ contain all of the nearest neighbour exchange interactions on type $A$ or type $B$ tetrahedra, respectively, and 
are related to each other by half of the space group symmetry operations. 
A breathing mode distortion removes the symmetry operations relating $H_A$ and $H_B$ (reducing the space group from $Fd4\bar{m}$ to $F\bar{4}3m$), and instead of only four independent exchange constants there are now eight, four for each type of tetrahedron.  
Therefore, the Hamiltonian can be written as \cite{Curnoe_physrevb_2008}
\begin{eqnarray}
    H_{\rm breathe} 
& = & {\cal J}_{A1}X_{A1}+{\cal J}_{A2}X_{A2}+{\cal J}_{A3}X_{A3}+{\cal J}_{A4}X_{A4} \nonumber \\
&  +&  {\cal J}_{B1}X_{B1}+{\cal J}_{B2}X_{B2}+{\cal J}_{B3}X_{B3}+{\cal J}_{B4}X_{B4}.
\label{distortion ham}
\end{eqnarray}
In an extreme limit, all 
of the constants associated with one orientation of tetrahedron (say $B$) will vanish, and eigenstates of 
$H_{\rm breathe}$ will be tensor products of single tetrahedron eigenstates. In this case, there is only one possible singlet state, 
$|A_1\rangle \otimes |A_1\rangle \otimes
|A_1\rangle \otimes
|A_1\rangle$.

The complete spectrum of $H$ and $H_{\rm breathe}$ 
for a 16-site cluster was obtained by exact diagonalization for a range exchange constants in the vicinity of the quantum spin ice phase \cite{Wei_2023, wei2}. In the following, we use the singlet ground states obtained from these calculations. 


\section{Results}

\subsection{${\cal I}$-Concurrence of singlet states (no lattice distortion)}
\begin{figure}[ht]
\centering
\subfloat{
  \includegraphics[width=27mm]{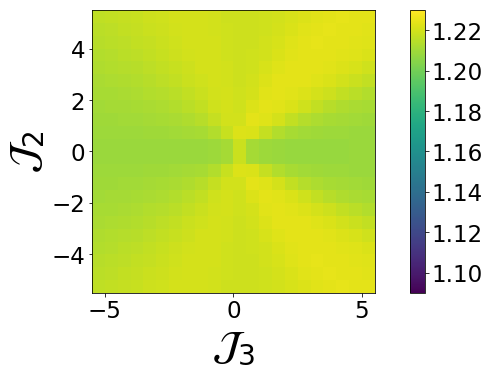}
}
\subfloat{
  \includegraphics[width=27mm]{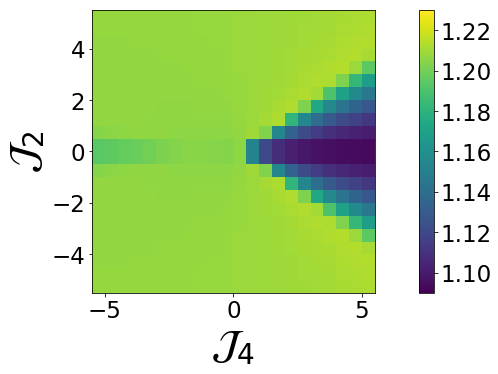}
}
\subfloat{
  \includegraphics[width=27mm]{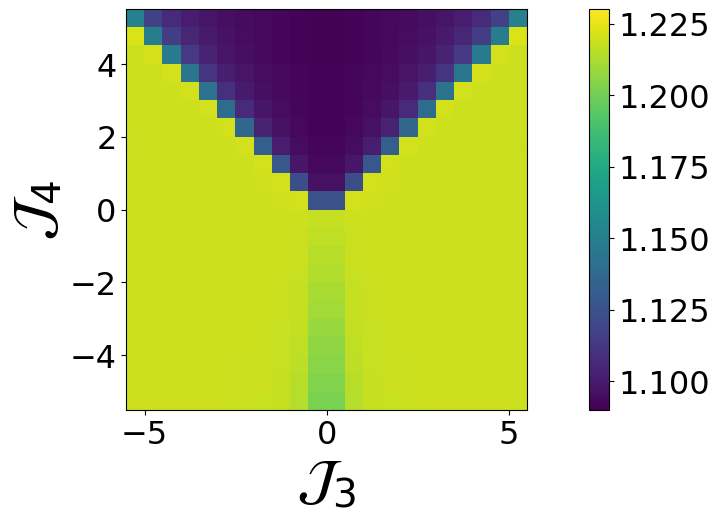}
}
\hspace{0mm}
\subfloat{
  \includegraphics[width=27mm]{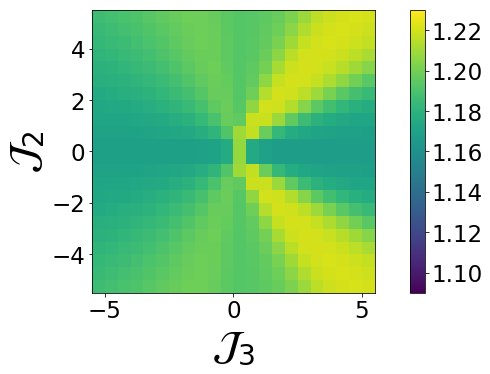}
}
\subfloat{
  \includegraphics[width=27mm]{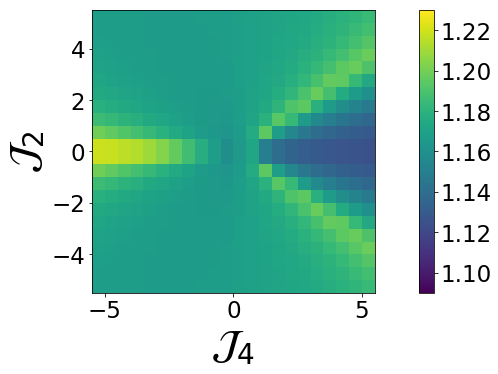}
}
\subfloat{
  \includegraphics[width=27mm]{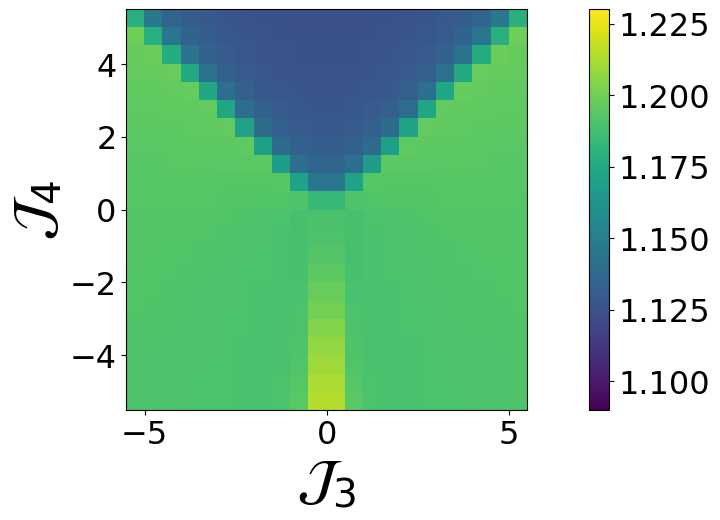}
}
\hspace{0mm}
\subfloat{
  \includegraphics[width=27mm]{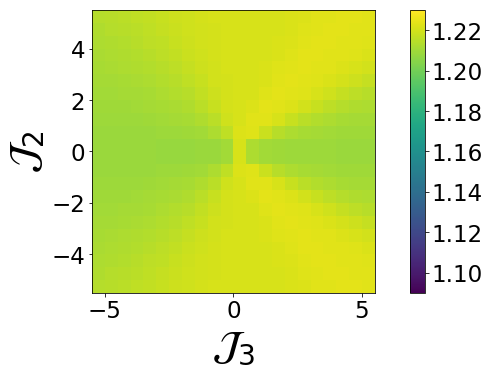}
}
\subfloat{
  \includegraphics[width=27mm]{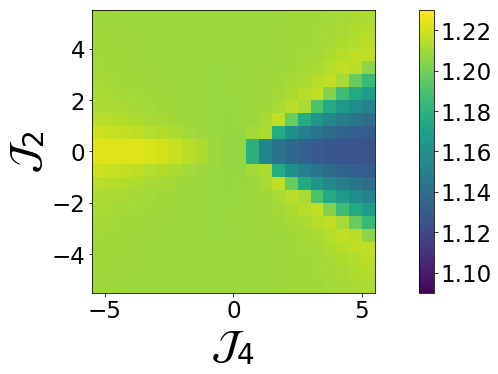}
}
\subfloat{
  \includegraphics[width=27mm]{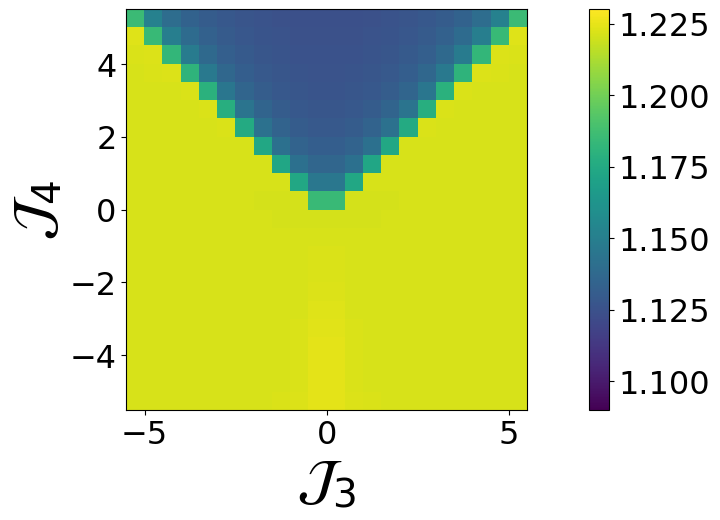}
}

\caption{2-site ${\mathcal {I}}$-concurrence of the ground state in the vicinity of the spin ice phase (${\cal J}_1 =-1$). Top row: nearest neighbours; middle row: next-nearest neighbours; bottom row: next-next-nearest neighbours. 
\label{fig:16-3}}
\end{figure}

In the $N=16$-site pyrochlore spin system, there are many different ways to define ${\cal I}$-concurrence according to which subset of spins comprises the subsystem used in calculating (\ref{eq:Iconcur}). In this paper we examine the following choices: {\em i)}
a single spin; {\em ii)} two spins that are 
{\em a)} nearest neighbours, {\em b)} next-nearest neighbours, or {\em c)} next-next-nearest neighbours; and {\em iii)} four spins on a single tetrahedron.
\\
\\
\noindent\underline{Single spin (1-site):} As long as there is a singlet ground state, the 1-site
{\cal I}-concurrence is constant and equal to the maximum value of one for all range of parameters. \\
\\
\noindent\underline{Two spins (2-site):}
The 2-site ${\cal I}$-concurrence for a singlet ground state is shown in Fig.\ \ref{fig:16-3}.  The 2-site subsystem has
dimension four, and so the maximum ${\cal I}$-concurrence is $\sqrt{3/2} = 1.2247$.  In our 16-spin system with periodic boundary conditions, all pairs of spins are either nearest neighbours, next-nearest neighbours or next-next-nearest neighbours; each spin has six nearest-neighbours (as they would in an infinite crystal), three next-nearest neighbours and six next-nearest neighbours.  Since each spin occupies a vertex shared between an $A$ tetrahedron and a $B$ tetrahedron, its six nearest neighbours are the spins occupying the other vertices of both tetrahedra.  If the spin in question is at a \#1 site, its nearest-neighbours are at \#2, 3 and 4 sites (two of each), its next-nearest neighbours are at \#1 sites, and the next-next-nearest neighbours are at \#2, 3 and 4 sites (two of each).
The results for each of these possibilities are shown in each of the three rows of Fig.\ \ref{fig:16-3}. 

\begin{figure}[ht]
\centering
\subfloat{
  \includegraphics[width=27mm]{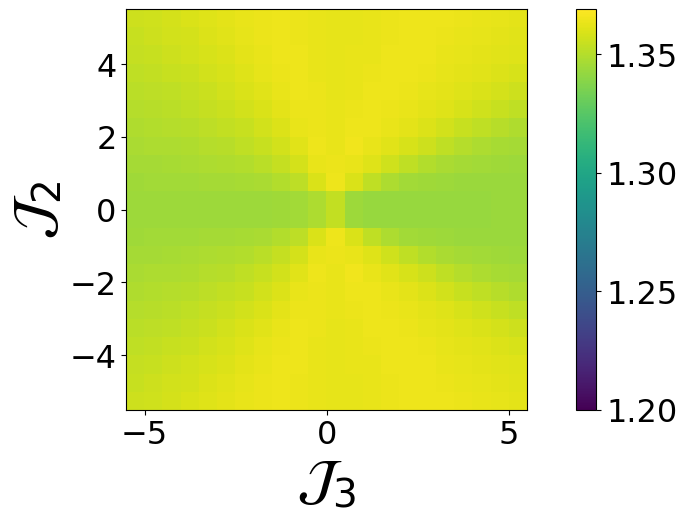}
}
\subfloat{
  \includegraphics[width=27mm]{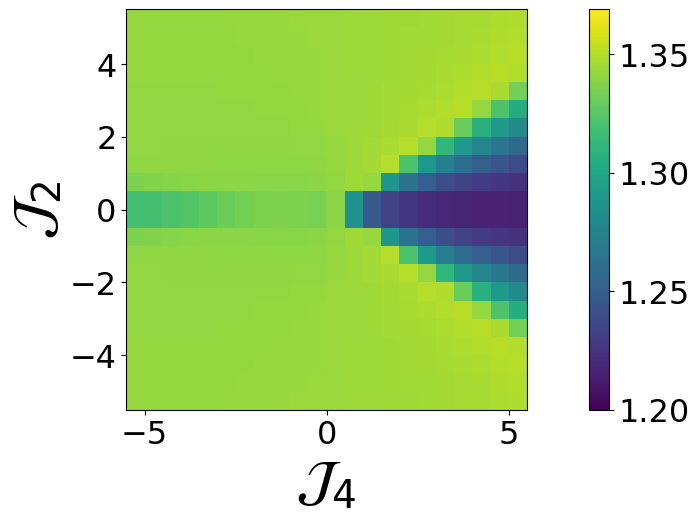}
}
\subfloat{
  \includegraphics[width=27mm]{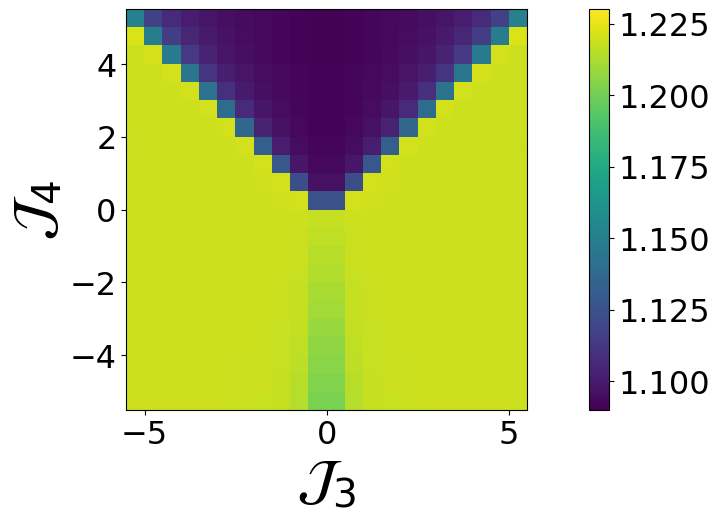}
}
\caption{1-tetrahedron ${\mathcal {I}}$-Concurrence of the ground state in the vicinity of the spin ice phase (${\cal J}_1 = -1$). 
\label{fig:16-2}}
\end{figure}

All of the plots show variations in the ${\cal I}$-concurrence, but these are especially pronounced as a function of the exchange parameter ${\cal J}_4$.  
The term $X_4$ raises or lowers adjacent spins in a tetrahedron, preserving the ice-rule in that tetrahedron but yielding
`3-in-1-out' or `1-in-3-out' states in adjoining tetrahedra. The sharp features appearing in Fig. \ref{fig:16-3} are in fact continuous, and coincide with an avoided level crossing of a singlet state that is a superposition of spin ice states and a singlet state that contains 3-in-1-out/1-in-3-out states \cite{Wei_2023}. In a previous work, we calculated the density of 2-in-2-out, 3-in-1-out/3-out-1-in
and all-in/all-out configurations as a function of the coupling constants ${\cal J}_a$; 
{\em all} of the radial features appearing on the plots in Figs.\ \ref{fig:16-3} and \ref{fig:16-2} correspond to crossovers between states with 
different characteristics observed in our previous work \cite{Wei_2023}.

The ${\cal I}$-concurrence for next-nearest neighbours is markedly smaller than for either nearest neighbours or next-next-nearest neighbours. Similar to the ${\cal I}$-concurrence of two spins in a single tetrahedron $|A_1\rangle$ state (see Section \ref{tetra}), some terms of the 16-site spin ice state can be factored; there are more of them when 
next-nearest neighbours are selected as the subsystem.  
\\

\noindent\underline{Four spins (1-tetrahedron):}
Here we consider a 4-site ${\cal I}$-concurrence, where the four sites are nearest neighbours with each other, shown in Fig.\ \ref{fig:16-2}. With $d=16$, the maximum concurrence is 
$\sqrt{15/8}= 1.3693$.  The results are very similar to the 
2-site (nearest-neighbour) ${\cal I}$-concurrence,
which is a reflects that 
a tetrahedron contains pairs of nearest neighbours.

\subsection{${\cal I}$-Concurrence with Lattice Distortion}
\begin{figure}[ht]
\centering
\subfloat[1-site ${\cal I}$-concurrence.]{
  \includegraphics[width=40mm]{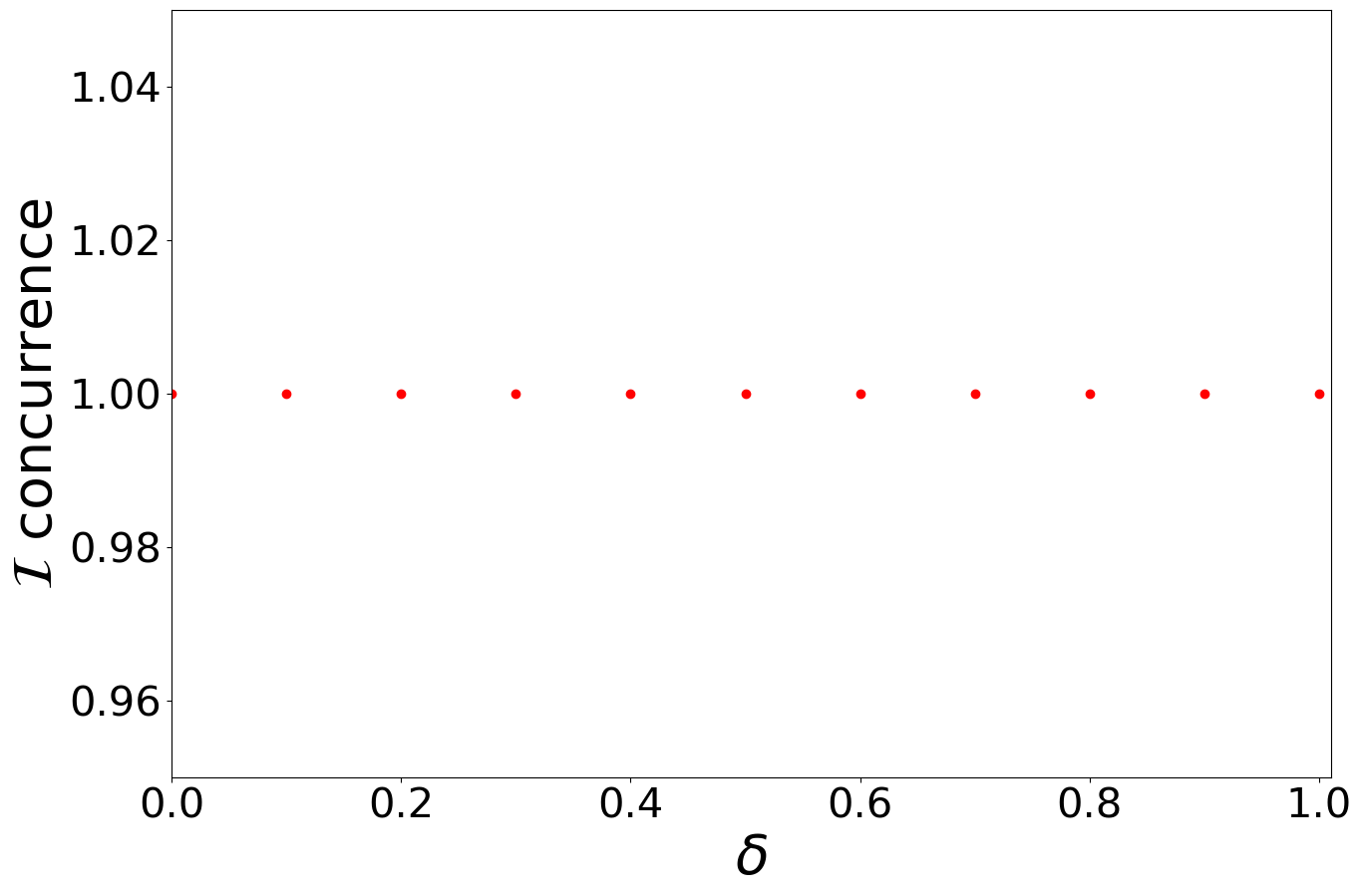}
}
\subfloat[1-tetrahedron ${\cal I}$-concurrence.]{
  \includegraphics[width=40mm]{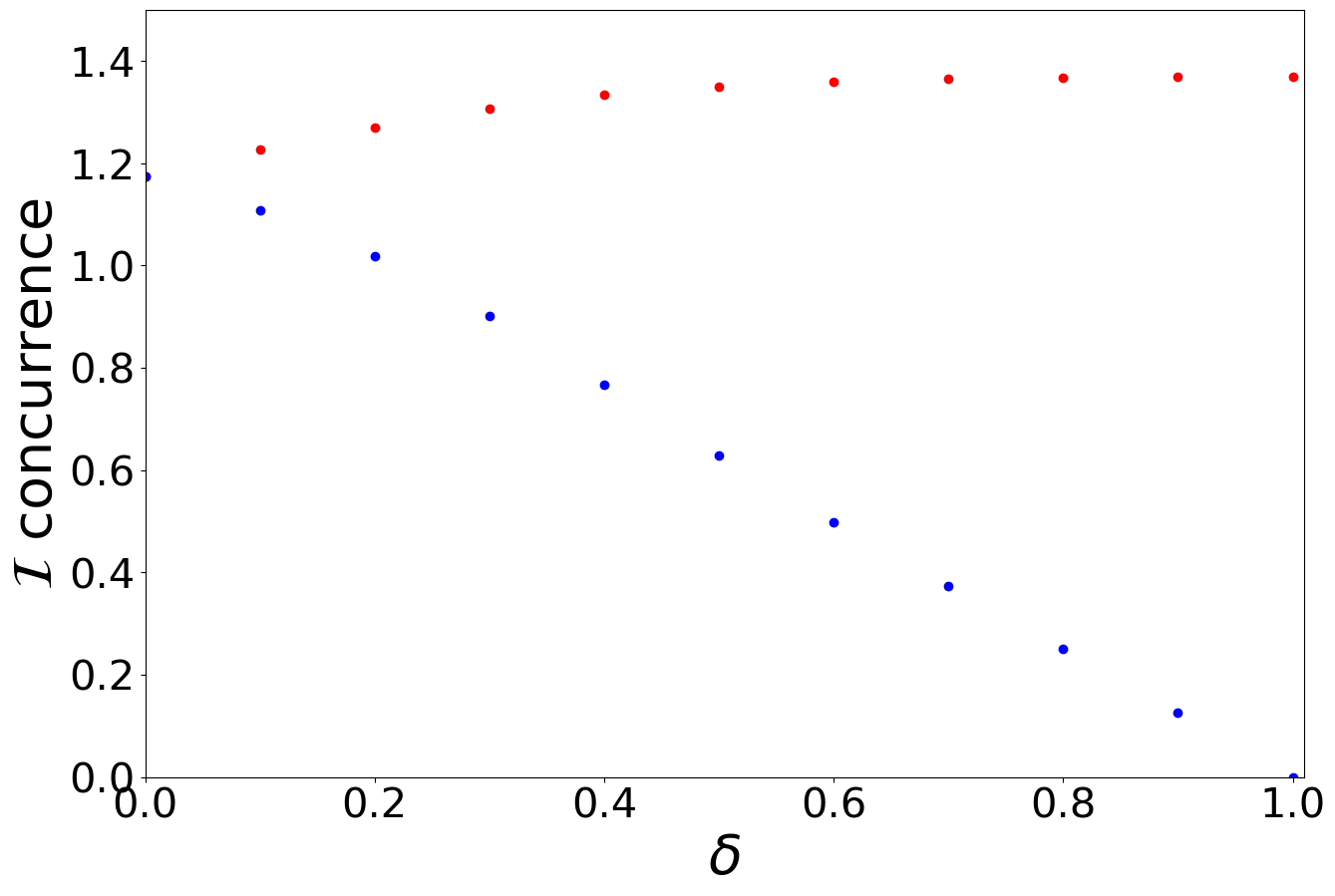}
}
\hspace{0mm}
\subfloat[2-site (nearest neighbour) ${\cal I}$-concurrence. Red: nearest neighbours on an $A$-tetrahedron; blue: nearest-neighbours on a $B$-tetrahedron.]{
  \includegraphics[width=40mm]{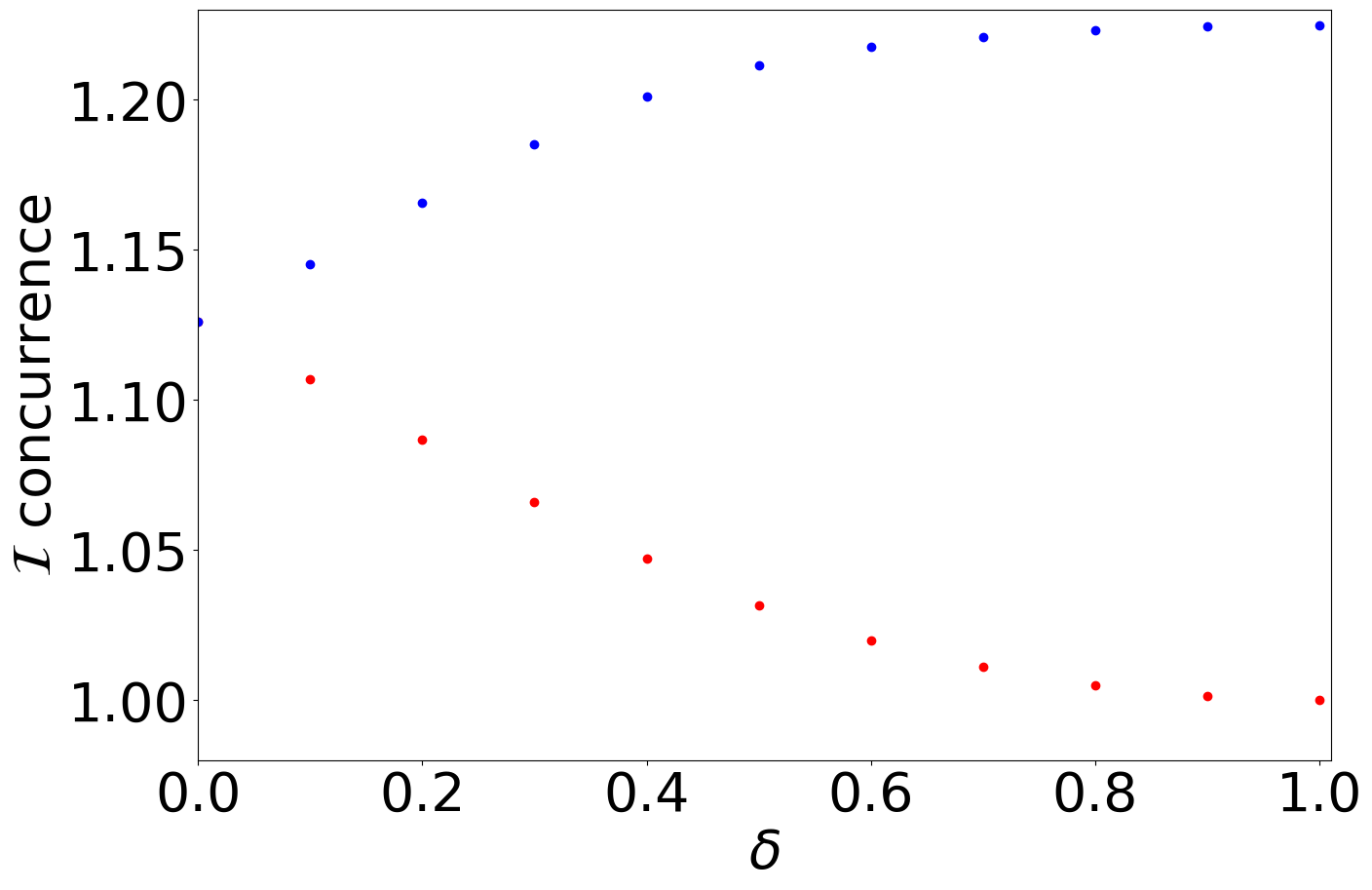}
}
\subfloat[2-site (next-nearest neighbour(blue) and next-next-nearest neighbour (red)) ${\cal I}$-concurrence.]{
  \includegraphics[width=40mm]{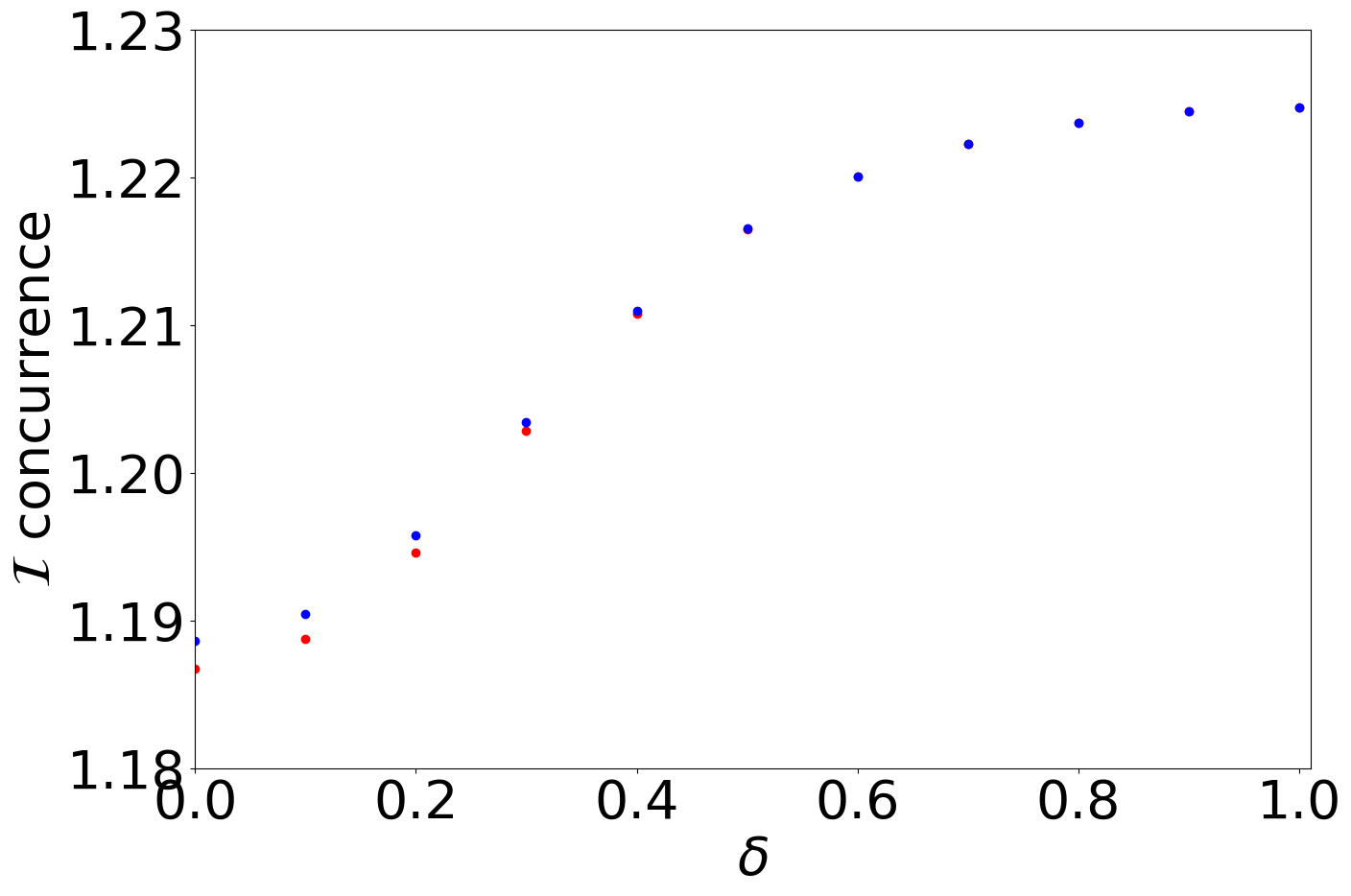}
}
\caption{Various ${\mathcal {I}}$-concurrences of the ground state as a function of the lattice distortion parameter
$\delta$. 
\label{fig:16-4}}
\end{figure}

In order to better understand the results of the previous section, 
we also consider the effect of a lattice distortion by varying the coupling constants on the $A$ and $B$ tetrahedra.  Fig.\ \ref{fig:16-4} shows the 1-site, 2-site and 1-tetrahedron ${\cal I}$-concurrences calculated using the ground state obtained from exact diagonalization. The lattice distortion is implemented by the parameter $\delta$: $\delta = 0$ represents no distortion and $\delta = 1$ results in a lattice where the exchange constants on the $B$ tetrahedra vanish and the $A$ tetrahedra are completely decoupled from each other.  The exchange constants used in the simulation are
${\cal J}_{Ai}={\cal J}_i(1+\delta)$ 
and ${\cal J}_{Bi}={\cal J}_i(1-\delta)$ where we have used 
${\cal J}_{1} = -1$,
${\cal J}_2 = 0.1$,
${\cal J}_3 = 0.2$, and
${\cal J}_4 = 0.3$.

Fig.\ \ref{fig:16-4} a) shows the 1-site ${\cal I}$-concurrence, which remains constant at the maximum value of one for any amount of distortion. 
Fig.\ \ref{fig:16-4} b) shows the 1-tetrahedron ${\cal I}$-concurrence for $A$ and $B$ tetrahedra.  The ${\cal I}$-concurrence for a $B$ tetrahedron increases with the amount of distortion and approaches maximum value, but for an $A$ tetrahedron it decreases and vanishes in the limit where the tetrahedra are completely decoupled. In this limit the state
is a product over $A$-tetrahedra, $|A_1\rangle \otimes |A_1\rangle \otimes|A_1\rangle \otimes|A_1\rangle$, which is separable for the $A$-tetrahedra but cannot be re-expressed as a simple product over $B$-tetrahedra.

Fig.\ \ref{fig:16-4} (c) shows the 2-site ${\cal I}$-concurrence for nearest neighbours.  There are two possibilities: either the nearest-neighbours are found on an $A$-tetrahedron or they are found on a $B$-tetrahedron. If they are on an $A$-tetrahedron then in the limit of large distortion, they will only be entangled with the other two spins on the $A$-tetrahedron (see Section \ref{tetra}).  The ${\cal I}$-concurrence does not vanish, but it does decrease to a value of 1, which is expected because this is the value of the ${\cal I}$-concurrence of two spins in a single tetrahedron.
If the neighbouring spins are on a $B$-tetrahedron then in the limit of a large distortion they will appear two different factors associated with the two $A$-tetrahedra they reside on, and the ${\cal I}$-concurrence will reflect each spin's entanglement with the other spins on its $A$-tetrahedra, which will be a maximum.  Similarly, the ${\cal I}$-concurrence for any other pair of spins that are not nearest-neighbours on a $A$-tetrahedron will approach the maximum value in the limit of a large distortion, as shown in Fig.\ \ref{fig:16-4} d).

To summarize, the effect of a lattice distortion is to factorize the state into factors associated with a set of tetrahedra, which tends to reduce the entanglement between tetrahedra. 
Hence the 
1-tetrahedron ${\cal I}$-concurrence, which measures entanglement between a tetrahedron and the rest of the system, vanishes in the limit of a large distortion factor. the disentanglement of a tetrahedron from the rest of the system is also evident in reduction of the 2-site ${\cal I}$-concurrence when the 2-sites are on  the tetrahedron. 
In general, the ${\cal I}$-concurrence for smaller subparts, such as 1-site, or 2-site, will always be non-vanishing because of entanglement with other parts {\em within} a tetrahedron; the disentanglement of a tetrahedron from the rest of the system is also evident in the 2-site ${\cal I}$-concurrence when the 2-sites are found together on a disentangled tetrahedron.


\section{Discussion}


We have calculated the ${\cal I}$-concurrence for a pyrochlore spin system - a multipartite, 2D system on a highly symmetric lattice - to evaluate the entanglement between different sized subsystems. 
The symmetry of the system allows for multifold degeneracies of the spin eigenstates, while also permitting a number of singlet states; for a limited range of model parameters a singlet will be the ground state \cite{note1}.  All of the singlet eigenstates must transform into themselves under the symmetry operations of the crystal, which means that generally they will be 
superpositions of many classical spin states.  In the finite cluster that we have considered, symmetry ensures that a state of the form 
$| \alpha \beta \gamma \ldots\rangle $ will always occur with an equal amplitude superposition of the state
$|\bar{\alpha} \bar {\beta} \bar{\gamma} \ldots\rangle$.  However, the 
state $|\alpha \beta \gamma\ldots\rangle$ 
will never appear in a superposition 
with $| \bar{\alpha} \beta \gamma \ldots\rangle$,
also because of symmetry:
there are no matrix elements of the Hamiltonian (to any order)
connecting them.
Therefore the reduced density matrix for any single spin will always be $\left(\begin{array}{cc}
1/2 & 0 \\
0 & 1/2 \end{array}\right)$, yielding the maximum 1-site ${\cal I}$-concurrence of one. Note that this is also guaranteed by time-reversal symmetry if there are an even number of spins, and is upheld even when there is a breathing mode lattice distortion. However, the loss of the three-fold axes (as in a cubic to tetragonal transition) will cause a reduction of the 1-site ${\cal I}$-concurrence. 

While a singlet ground state does not occur in all regions of the parameter space spanned by exchange constants ${\cal J}_i$, it does occur in one region of special interest,  ${\cal J}_1 < 0$ and
$|{\cal J}_{2,3,4}| \lesssim |{\cal J}_1|$, which is the expected location of a 
 {\em quantum spin ice} phase, and is where the exchange constants of several real quantum spin ice candidates, including 
 Tb$_2$Ti$_2$O$_7$ \cite{Gardne_physrevb_2001, curnoe_2013_PhysRevB.88.014429}, Yb$_2$Ti$_2$O$_7$ \cite{Ross_physrevX_2011,chang_2021_nature_commun, pan_nature_2014, robert_2015_PhysRevB.92.064425}, 
and several Pr oxides 
\cite{zhou_2008_PhysRevLett.101.227204,sibille_2016_PhysRevB.94.024436, petit_2016_PhysRevB.94.165153}, are found. 
At the point at the very centre  of the plots, the classical states which minimize the energy are those that obey the ice rule: two spins pointing into and two spins pointing out of each tetrahedron. The number of these states scales with the system size as approximately  $1.5^{N/2}$ (where $N$ is the number of spins) \cite{pauling1935}, and similarly to water ice, produces a low temperature residual entropy. It is important to note that these states are degenerate at the point 
$|{\cal J}_{2,3,4}| = 0$; here the system is described as a density matrix rather than a pure state, 
and all the various concurrence measures vanish. The lowest energy excitations are  ice rule violations on single tetrahedra.  

When the exchange constants $|{\cal J}_{2,3,4}|$ are non-zero there will be 
perturbative lifting of the spin ice degeneracy and mixing of excited states into the ground state \cite{Wei_2023}.  Nevertheless, there is a well-defined, finite region in the vicinity of ${\cal J}_1 = -1$, 
$|{\cal J}_{2}| \lesssim 0.1 $, and 
$|{\cal J}_{3,4}| \lesssim 0.25$
 where the ice rule is largely obeyed and where a quantum spin ice phase is expected to be found \cite{Wei_2023}.  This region is actually a very small area at the  centre of the plots shown in Figs.\ \ref{fig:16-3} and \ref{fig:16-2},  where the
 concurrence of the ground state singlet is large.  However, while the ice rule is obeyed only in the centre of the region shown in our plots, entanglement, as quantified by the 
 ${\cal I}$-concurrence, persists well beyond this region.

\subsection*{Summary}\label{sec:Summary}
The ${\cal I}$-concurrence measures the entanglement between two complementary {\em subsystems} of system.  It does not directly measure the entanglement between any two parts, except in simple bi-partite systems, or when the two parts are the two subsystems under consideration, but it does reflect the parameter-dependent composition of the state as well as
the underlying symmetry of the system. 

Using the ${\cal I}$-concurrence as an entanglement measure, we have studied the entanglement of the singlet ground state of a 
 16-site pyrochlore cluster using results obtained from exact diagonalization on the four parameter nearest-neighbour exchange Hamiltonian.  We focused on a region of the parameter space where there is a singlet ground state and which contains the smaller region where a quantum spin ice phase is expected.  
 Singlet states are necessarily (because of symmetry) superpositions of large numbers of classical states, and so entanglement, as quantified by the 
 ${\cal I}$-concurrence, is large throughout this region.
Variations in the ${\cal I}$-concurrence reflect variations the composition of the ground state. The ${\cal I}$-concurrence also reveals how a symmetry-lowering lattice distortion disentangles the ground state.

\begin{acknowledgments}
This work was supported by the Natural Sciences and Engineering Research Council of Canada.
	\end{acknowledgments}

	\appendix





\bibliographystyle{apsrev4-1}
    \bibliography{citations}

\end{document}